\documentclass[aps,prd,twocolumn,nofootinbib,floatfix,showpacs,showkeys,preprintnumbers]{revtex4}
\usepackage{subfigure}
\usepackage{graphicx}
\usepackage{dcolumn}
\usepackage{epsfig}
\usepackage{amsmath}
\usepackage{amsfonts}
\usepackage{amssymb}
\usepackage{color}
\usepackage{hyperref}

\begin{document}

\title{Cosmic Reionization History and Dark Matter Scenarios}

\author{Isabella P. Carucci}
\email{i.carucci@ucl.ac.uk}
\affiliation{Department of Physics and Astronomy, University College London, London WC1E 6BT, UK}

\author{Pier-Stefano Corasaniti}
\affiliation{LUTH, UMR 8102 CNRS, Observatoire de Paris, PSL Research University, Universit\'e Paris Diderot, 92190 Meudon -- France}

\begin{abstract}
We perform an analysis of the cosmic reionization in the standard cold dark matter (CDM) paradigm and in alternative dark matter scenarios. Building upon the work of Corasaniti et al.~(2017), we predict the reionization history for CDM, for warm dark matter (WDM), late-forming dark matter (LFDM) and ultra-light axion dark matter (ADM) models which reproduce state-of-art measurements of the galaxy luminosity function at very high-redshifts $6\le z\le 10$. To this purpose we adopt a reionization model parametrized in terms of the limiting UV-magnitude of galaxies contributing to the reionization $M_{\rm lim}$ and the average effective escape fraction of UV photons reaching the intergalactic medium $\tilde{f}$. For each DM model we compute the redshift evolution of the Thomson scattering optical depth $\tau_e(z)$ and the comoving ionization fraction $Q_{\rm HII}(z)$. We find the different DM models to have similar reionization histories. Differences with respect to the CDM case increase at fainter limiting UV-magnitudes and are degenerate with the effect of varying the reionization model parameters. Using {\it Planck}'s determination of the integrated optical depth in combination with measurements of the neutral hydrogen fraction at different redshifts, we infer constraints on $\tilde{f}$ and $M_{\rm lim}$. The results are largely independent of the assumed DM scenario, in particular for $M_{\rm lim}\gtrsim -13$ we obtain that the effective escape fraction lies in the range $0.07\lesssim \tilde{f}\lesssim 0.15$ at $2\sigma$.

\end{abstract}


\maketitle

\newcommand{\ste}[1]{\textcolor{red}{\textbf{\small[Ste: #1]}}}
\newcommand{\isa}[1]{\textcolor{blue}{\textbf{\small[isa: #1]}}}

\section{Introduction}\label{intro}
The physical mechanisms that drive the ionization of the neutral hydrogen in the universe have yet to be fully understood. Nevertheless, over the past years cosmological observations have opened new windows of investigation on the cosmic reionization history. As an example, the EDGES experiment has recently claimed the detection of a 21-cm absorption signal \cite{EDGES} indicating that star formation must have produced ionizing radiation by redshift $z\approx 20$. Although the interpretation of the signal is still being debated, unambiguous measurements of the temperature and polarization anisotropies of the Cosmic Microwave Background (CMB) radiation from the {\it Planck} satellite \cite{Planck2016,Planck2018} have provided information on the late-time phase of the cosmic reionization process. In particular, the bounds on integrated Thomson scattering optical depth \cite{PlanckCosmo2018} suggests that the reionization process is completed by redshift $z\approx 7$. Yet, we are far from having acquired a comprehensive picture of how this epoch has evolved, as pointed out by observations of quasars spectra \cite{Bosman2018}. 

There is a consensus that future 21-cm observations from the present and upcoming radio facilities such as LOFAR\footnote{\url{http://www.lofar.org}}, MWA\footnote{\url{http://mwatelescope.org}}, PAPER\footnote{\url{http://eor.berkeley.edu}}, GMRT\footnote{\url{http://www.gmrt.ncra.tifr.res.in}}, HERA\footnote{\url{http://reionization.org}} and SKA\footnote{\url{https://www.skatelescope.org}} may provide a detailed picture of the reionization. However, progress in this direction has also come in recent years from the optical detection of very faint galaxies at $z\gtrsim 6$. The realization of observational programs such as the Hubble Ultra Deep Field \cite{HUDF} and Hubble Frontier Field \cite{HFF} have allowed unprecedented measurements of the abundance of very faint high-redshift galaxies that are believed to be the primary source of ionizing UV-radiation (see e.g.~\cite{Bouwens2015,Finkelstein2015,Atek2015a,Atek2015b,Livermore2017,Bouwens2017}). These measurements have sparked ample studies of the relation between early galaxy formation and cosmic reionization (see e.g. \cite{Jaacks2012,Kuhlen2012,Oshea2015,Mason2015,Mashian2016,Liu2016,Ishigaki2018}), but have far wider implications since they also indirectly probe the nature of dark matter (DM) in the universe.

In the standard cosmological model, DM consists of cold collisionless particles interacting with baryonic matter through gravity only. Inspired by high-energy theories beyond the Standard Model of particle physics, this so called cold dark matter (CDM) paradigm has been tremendously successful at reproducing the observed distribution of cosmic structures on the large scales. However, the emergence of anomalies at small scales and the lack of detection of CDM particle candidates, such as weakly interacting massive particles (WIMPs) in physics laboratories, have motivated the study of alternative DM models in which DM particles evade direct detection. This is the case of sterile neutrinos (see e.g. \cite{WDMlit}) with a thermal relic particle mass of order a few keV, also referred as warm dark matter. Ultra-light axions and scalar field DM models have also been proposed in the literature as alternative DM particle candidates (see e.g. \cite{DoddyRev} for a review).

A distinct feature of these models is the suppression of the abundance of low mass DM halos which host the very faint galaxies recently observed in the high-$z$ universe. As the UV-radiation from these galaxies contribute to the cosmic reionization, we expect such non-standard DM scenarios to leave an imprint on the reionization history of the universe. 

Constraints on DM models from measurements of the galaxy luminosity function (LF) and their impact on the cosmic reionization history have been investigated in several works in the literature \cite{Schultz2014,Lapi2015,Bozek2015,Schive2016,Lopez2017}. A key point of these analyses concerns the specification of the relation between the DM halo mass and the host galaxy UV-luminosity. In principle, such a relation solely depends on the astrophysical processes responsible for the formation of galaxies. However, contrary to the assumptions of \cite{Schultz2014}, these mechanisms may occur differently depending on the underlying properties of DM, giving rise to a DM model dependent relation (see e.g. \cite{Lapi2015,Bozek2015,carucci2015,Schive2016,Corasaniti2017,Lopez2017}). As an example, the authors of \cite{Corasaniti2017} have shown that DM models with suppressed abundances of low mass halos can reproduce the observed luminosity functions provided that the star formation rate at low halo masses is higher than in the standard CDM scenario. This trend has also been confirmed in \cite{Dayal2017}, where the authors have used a semi-analytic approach to model the formation of high-redshift galaxies in WDM models, and in \cite{Villanueva2018} using numerical hydrodynamics simulations. This points to a degeneracy between the imprint of non-standard DM models on the galaxy luminosity function and the specifics of the galaxy formation process that should also affect the cosmic reionization history (see e.g. \cite{Lopez2017}). 

Here, we present a detailed study of the interplay between the imprint of non-standard DM models on the faint-end of the high-$z$ galaxy luminosity function and the cosmic reionization. 
The paper is organized as follows. In Section~\ref{intro} we describe the dark matter scenarios considered in the analysis, the analytical modeling of the cosmic reionization and the basic model assumptions. In Section~\ref{resultsI} and \ref{resultsII} we discuss the results of the computation and the comparison with existing bounds of the cosmic reionization history. In Section~\ref{MCMC} we show a complete assessment of the parameter space of the reionization model. Finally, in Section~\ref{conclu} we present our conclusions.

\section{Methodology}

\begin{figure} 
	\centering
	\includegraphics[width=1\hsize]{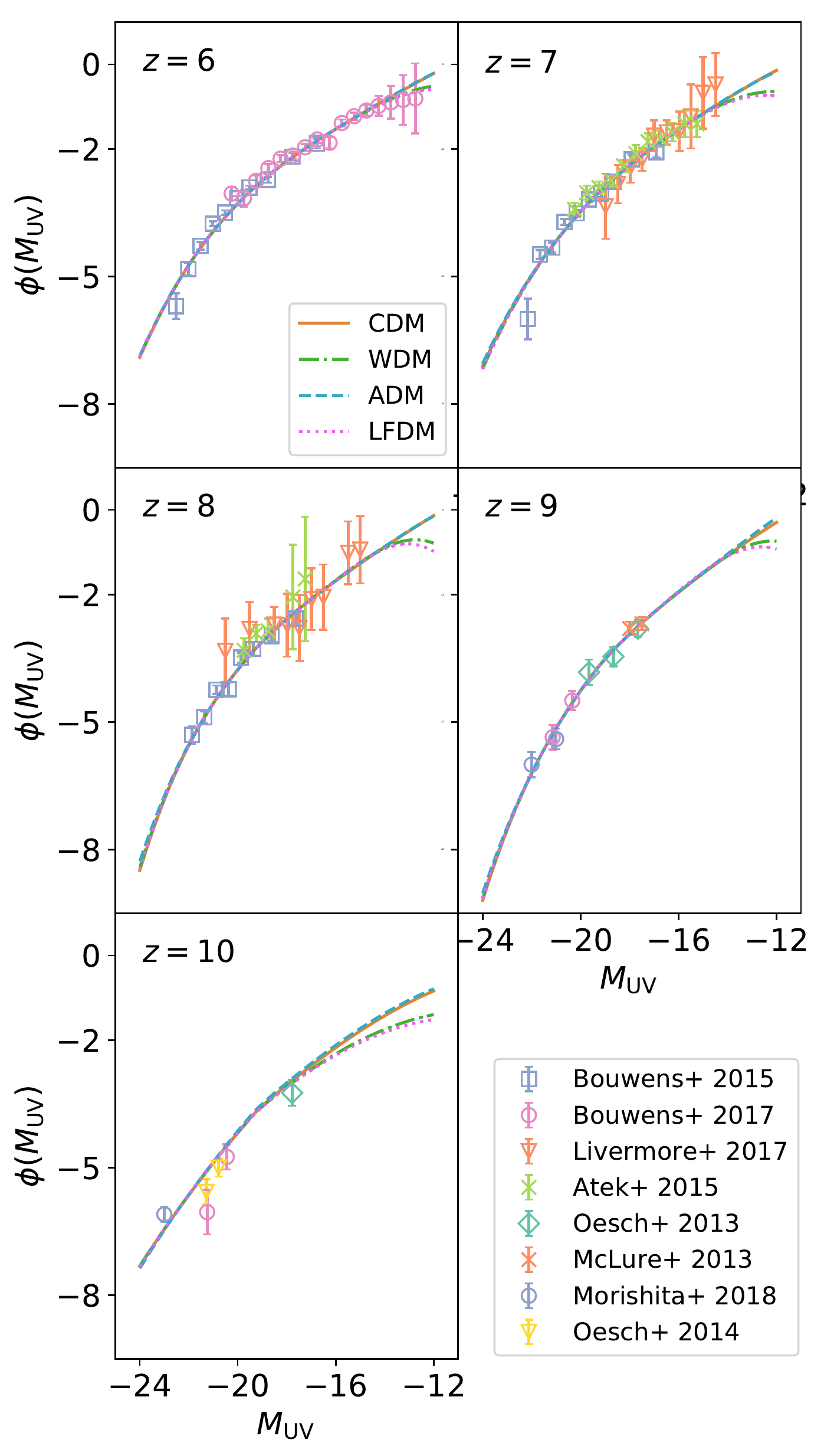}
	\caption{\label{fig:LF} Best-fit galaxy luminosity functions obtained at redshift $z=6$ (top left panel), $z=7$ (top right), $z=8$ (middle left), $z=9$ (middle right) and $z=10$ (bottom left). The best-fit curves correspond to the CDM model (solid lines), WDM (dot-dashed), ADM (dashed) and LFDM (dotted). The different data points corresponds to LF measurements from Bouwens et al. \cite{Bouwens2015} (blue squares) \cite{Bouwens2017} (pink circles), Livermore et al. \cite{Livermore2017} (orange triangles), Atek et al. \cite{Atek2015b} (green crosses), Oesch et al. \cite{Oesch2013} (cyan diamonds), McLure et al. \cite{McLure2013} (orange crosses), Morishita et al. \cite{Morishita2018} (blue circles) and Oesch et al. \cite{Oesch2014} (yellow triangles).}
\end{figure}

We consider the DM models investigated in \cite{Corasaniti2017} and derive predictions of the corresponding cosmic reionization history using the analytical formalism presented in \cite{Kuhlen2012}. Consistent with the study presented in \cite{Corasaniti2017}, hereafter we assume a background flat cosmological model with cosmological constant $\Lambda$ specified by the following set of parameter values: mean matter density $\Omega_m=0.3$, baryon density $\Omega_b=0.046$, reduced Hubble parameter $h=0.7$, scalar spectra index $n_s=0.99$ and root-mean-square fluctuation amplitude $\sigma_8=0.8$.

\subsection{Dark Matter Scenarios}
In \cite{Corasaniti2017}, the authors have inferred constraints on several DM models from state-of-art measurements of the high-redshift galaxy luminosity function. More specifically they have considered warm dark matter (WDM) models characterized by different values of the thermal relic particle mass, ultra-light axions (ADM) with different particle masses and late-forming scalar dark matter (LFDM, \cite{Das2011}) models with different transition redshifts. For these scenarios, they have realized a series of high resolution N-body simulations to accurately resolve the low mass end of the high redshift halo mass function. In order to convert halo masses into UV-magnitudes of the hosted galaxies, they have adopted a hybrid abundance matching method to derive model predictions of the galaxy LF which they have compared to a compilation of LF measurements at $z=6,7$ and $8$ from \cite{Atek2015b,Bouwens2015,Bouwens2017,Livermore2017}. Then, they have identified the best-fit models which reproduce the observed galaxy abundance at same statistical significance. These consist of a CDM model, a WDM model with thermal relic particle mass $m_{\rm WDM}=1.465$ KeV, an ADM model with axion mass\footnote{We note that the ADM model we adopt is currently challenged by the analysis of the Lyman-$\alpha$ forest power spectrum of \cite{irsic2017,nori2018} and by the analysis of rotation curves of well-resolved near-by galaxies of \cite{nitsan2018}. However, given the intrinsic complementarity of reionization with the latter probes  and the variety of assumptions of all them, we believe it is important to explore the ADM model to confirm or disprove the other constraining methods.}
$m_{a}=1.54\times 10^{-21}$ eV and a LFDM model with transition redshift $z_t=8\times 10^5$.

To perform a more accurate evaluation of the reionization history, we extend the validity of these models to also reproduce LF measurements at $z=9$ and $10$ from \cite{McLure2013,Oesch2013,Bouwens2016,Morishita2018,Oesch2014}. To this purpose, we follow the methodology developed in \cite{Corasaniti2017} which we briefly detail hereafter (further details are in Section 3 of \cite{Corasaniti2017}). First, we use the N-body calibrated mass functions from \cite{Corasaniti2017} to perform an abundance matching evaluation of the average UV-magnitude halo mass relation at $z=4$ and $5$ using the UV-LF \cite{Bouwens2015}. Then, we input these calibrated relations into a probabilistic model of the galaxy LF that for each DM model depends on the overall amplitude of the UV-magnitude halo mass relation and the scatter. Finally, we perform a Markov chain Monte Carlo analysis to determine the best-fit values of the LF model parameters and the DM model goodness-of-fit.

We limit the analysis to halo masses $M_{\rm h}>5\cdot~10^8\,M_{\odot}h^{-1}$, corresponding to the minimum halo mass resolved in the N-body simulations of \cite{Corasaniti2017}. It is possible that the minimum mass of star forming halos at these redshifts is below this value by $\sim1$ order of magnitude, corresponding to virial temperature for supporting atomic cooling. Nevertheless, other effects (e.g. supernova feedback, photo-heating from reionization) could disrupt star formation in these shallow potentials. We opt to be conservative and not to extrapolate below the halo mass resolution of our simulations.

\begin{figure}[t]
\centering
\includegraphics[width=1\hsize]{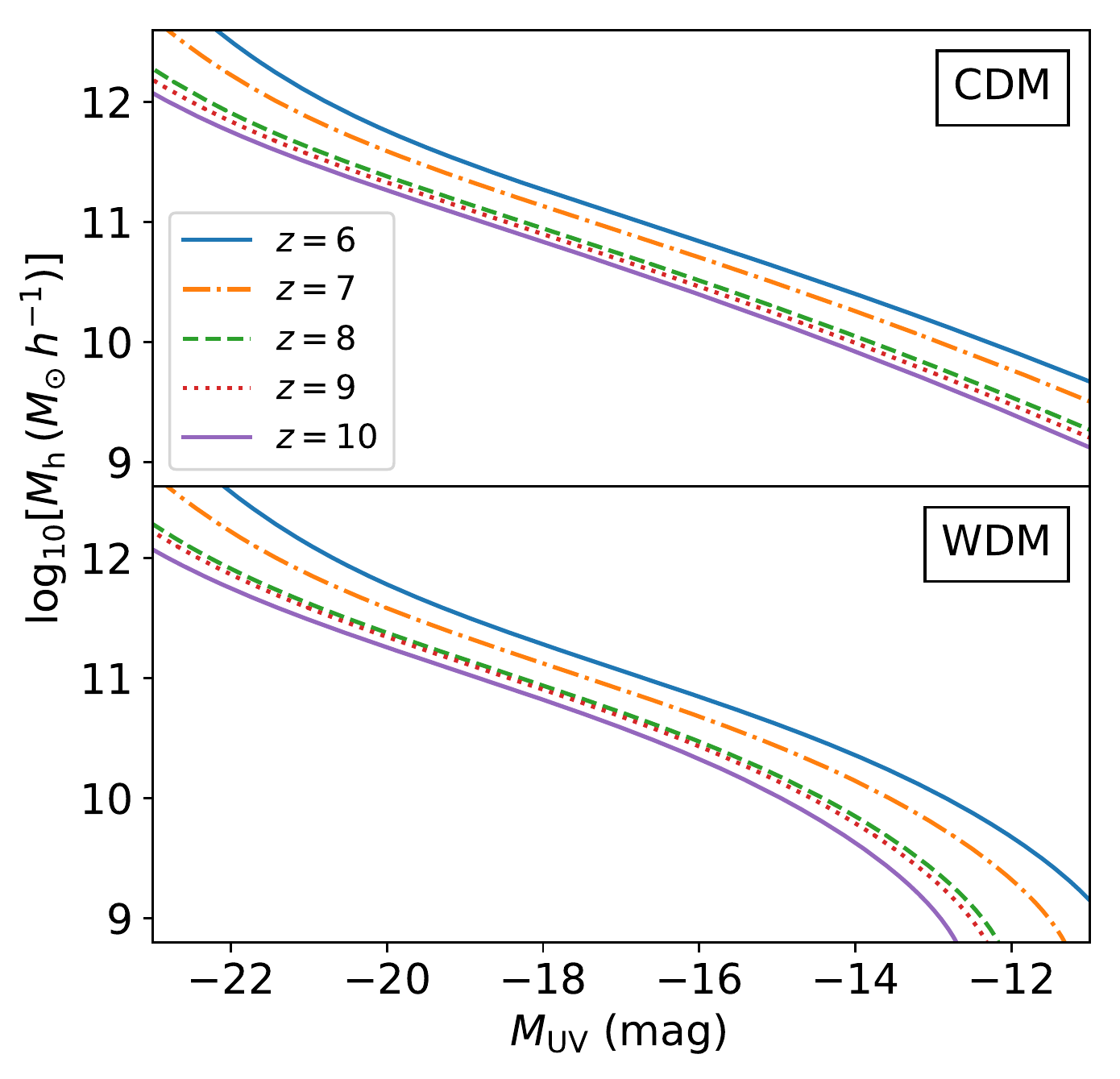}
\caption{\label{fig:MUVMH} Average UV-magnitude halo mass relation as inferred from best-fitting the galaxy LF function measurements at $z=6,7,8,9$ and $10$ (curves from top to bottom) for the CDM model (top panel) and WDM model (bottom panel).}
\end{figure}

In Fig.~\ref{fig:LF} we plot the best-fit galaxy LFs for the different DM models against the data at $z=6,7,8,9$ and $10$ in the UV-magnitude range $-24<M_{\rm UV}<-12$. We can see that the LF predicted by the ADM model shows no appreciable difference with respect to the CDM case, while WDM and LFDM exhibit a flattening of the faint-end slope at all redshifts. A consequence of these trends is that the different DM models predict different UV-magnitude halo mass relations. As an example, in Fig.~\ref{fig:MUVMH} we plot the average $M_{\rm UV}-M_{\rm h}$ relation for the CDM (top panel) and WDM (bottom panel) models at $z=6,7,8,9$ and $10$ (curves from top to bottom). 

Looking at Fig.~\ref{fig:MUVMH}, we may notice that while at the high-mass end the relations of the two models are identical, they differ at the low-mass end where the WDM scenario predicts brighter UV-magnitudes than the CDM case. Assuming that the UV-magnitude is an indicator of the star formation rate (SFR), this implies that in a WDM model the SFR is higher in low mass halos than in CDM. Such differences are necessary to compensate for the different low mass halo abundances of the two models and allow them to reproduce the observed LFs \footnote{It is important to remark that the relation between halo mass and UV-magnitude (or SFR) of the host galaxy adopted in the modeling of the galaxy LF is stochastic, characterized by a mean normalization amplitude and a scatter (see Eqs. 12-13 in \cite{Corasaniti2017}). Consistent with the results of \cite{Corasaniti2017,Mashian2016}, we find the scatter on the ${\rm SFR}-M_{\rm h}$ relation to be $\sim 0.5$ dex.}. 

\subsection{Cosmic Reionization Model} \label{sub:theory}
Here, we briefly review the analytical model of cosmic reionization presented in \cite{Kuhlen2012}. The rate of variation of the volume filling fraction of ionized hydrogen $Q_{\rm HII}(z)$ is given by
\begin{equation}\label{QHII}
\frac{dQ_{\rm HII}}{dt}=\frac{\dot{n}_{\rm ion}}{\bar{n}_{\rm H}}-\frac{Q_{\rm HII}}{\bar{t}_{\rm rec}},
\end{equation}
where $\dot{n}_{\rm ion}$ is the comoving rate of ionizing photon emission, $\bar{n}_{\rm H}$ is the mean comoving hydrogen number density\footnote{$\bar{n}_{\rm H}=0.75\Omega_b\rho_c/m_p$, where $\Omega_b$ is the cosmic baryon density, $\rho_c$ is the critical density and $m_p$ the proton mass.} and $\bar{t}_{\rm rec}$ is the volume average recombination time scale.
 
We assume that galaxies are the main source of ionizing radiation, as pointed out in e.g.  \cite{Robertson2015,Onorbe2017}. Then, the rate at which ionizing photons are emitted is given by
\begin{equation}
\dot{n}_{\rm ion}^{\rm com}=\int_{M_{\rm lim}}^{\infty} \phi(M_{\rm UV})\gamma_{\rm ion}(M_{\rm UV})f_{\rm esc}dM_{\rm UV},\label{ndotion}
\end{equation}
where $\phi(M_{\rm UV})$ is the galaxy luminosity function, $M_{\rm lim}$ is the limiting UV-magnitude, $f_{\rm esc}$ is the escape fraction of photons injected into the intergalactic medium (IGM) and $\gamma_{\rm ion}(M_{\rm UV})$ is the ionizing luminosity per UV-magnitude that depends on the galaxy spectral energy distribution (SED). In the case a double-power law model of SED, we have (see \cite{Kuhlen2012})
\begin{equation}
\gamma_{\rm ion}(M_{\rm UV})=2\times 10^{25}{\rm s}^{-1}\,\, 10^{0.4(51.63-M_{\rm UV})}  \zeta_{\rm ion},
\end{equation}
where $\zeta_{\rm ion}$ is a normalization which depends on the SED model parameters. 

The volume average recombination time scale reads as
\begin{equation}
\bar{t}_{\rm rec}\approx 0.93\,{\rm Gyr}\left(\frac{C_{\rm HII}}{3}\right)^{-1}\left(\frac{T_0}{2\times 10^4\,{\rm K}}\right)^{0.7}\left(\frac{1+z}{7}\right)^{-3},
\end{equation}
where $C_{\rm HII}$ is the effective clumping factor in ionized gas and $T_0$ is the temperature of the IGM at mean density. 

Here, we assume $T_0=2\times 10^{4}$ K, which is consistent with the results from IGM studies (see e.g. \cite{Schaye2000,Hui2003}). In contrast, the clumping factor is highly uncertain. Numerical simulations indicate a value in the range $2\lesssim C_{\rm HII}\lesssim 4$ (see e.g. \cite{Pawlik2009,Finlator2012}). Thus, consistent with previous works (e.g. \cite{Kuhlen2012,Schultz2014,Bozek2015}) and in agreement with the results of \cite{Gorce2017}, we set $C_{\rm HII}=3$.

The reionization model defined by Eqs.~(\ref{QHII})-(\ref{ndotion}) depends on $M_{\rm lim}$, $f_{\rm esc}$ and $\zeta_{\rm ion}$. These quantities parametrize astrophysical aspects of the reionization process that are not fully known. 

$M_{\rm lim}$ is the UV-magnitude of the faintest galaxy injecting ionizing photons in the IGM. Ultimately, this depends on the minimum star forming halo mass that may also vary in redshift. From Fig.~\ref{fig:MUVMH} we can see that a faint magnitude of $M_{\rm UV}=-12$ roughly corresponds to halos with mass $M_{h}\approx 10^9-10^{10}\, M_{\odot}\,h^{-1}$. Hence, in the following we estimate the reionization histories assuming $M_{\rm lim} \in [-16,-12]$. These values cover a conservative range of minimum star forming halo masses at high redshifts.

The escape fraction $f_{\rm esc}$ accounts for the fraction of photons which are not absorbed by dust or neutral hydrogen inside galaxies, and reach the IGM, thus contributing to the cosmic reionization process. This fraction may as well vary with galaxy properties and redshift. Indirect constraints have been inferred in \cite{Faisst2016} pointing to $f_{\rm esc}\approx 0.057\pm^{0.083}_{0.033}$ at $z=6$, while at the same redshift \cite{Kakiichi2018} obtain a lower bound $f_{\rm esc}\ge 0.08$. We summarize in Fig.~\ref{fig:fz} the available constraints on $f_{\rm esc}$ in the redshifts range $z \in [5.5, 9]$: these are consistent with a constant $f_{\rm esc}$ in this redshift range, however a redshift dependence cannot be excluded. We will discuss the case of a redshift dependent escape fraction in Section~\ref{resultsII}, where we evaluate the imprint of DM models on the cosmic reionization history assuming different redshift evolutions of $f_{\rm esc}$ consistent with such bounds. In the following, we focus on a constant escape fraction. In such a case $\dot{n}_{\rm ion}^{\rm com}$ is only sensitive to the product $\tilde{f}=f_{\rm esc}\zeta_{\rm ion}$, which we denote as effective escape fraction. Estimates of $\zeta_{\rm ion}$ from H$\alpha$ and UV-continuum fluxes of distant galaxies have been inferred in \cite{BouwensZetaIon2016}. Here, we assume values for $\tilde{f} \in [0.05,0.25]$ compatible with current observational constraints on $f_{\rm esc}$ and $\zeta_{\rm ion}$ respectively. 

In order to compute the cosmic reionization history of a given DM model we first determine $\dot{n}_{\rm ion}^{\rm com}$ for a set of values of $\tilde{f}$ and $M_{\rm lim}$ by computing Eq.~(\ref{ndotion}) assuming the best-fit LF of the assumed DM model at a given redshift\footnote{For simplicity we do not propagate LF uncertainties on the value of $\dot{n}_{\rm ion}^{\rm com}$  at a given redshift, since we extrapolate the trend to UV-magnitude above the currently available LF measurements.}. Then, we approximate the function $\dot{n}^{\rm com}_{\rm ion}(z)$ with a linear redshift interpolation over the redshift range $6\le z\le 10$ and impose a constant trend for $z<6$ with the value of the constant set to $\dot{n}^{\rm com}_{\rm ion}(z=6)$, while for $z>10$ we impose a cut-off $\dot{n}^{\rm com}_{\rm ion}=0$.  We have verified that the final results do not depend on such assumptions. Then, inserting  $\dot{n}^{\rm com}_{\rm ion}(z)$ in Eq.~(\ref{QHII}), we solve for $Q_{\rm HII}(z)$ with initial condition $Q_{\rm ini}=10^{-13}$ at $z_{\rm ini}=20$. We find a convergent solution for values of $Q_{\rm ini}$ that are negligibly small at $z_{\rm ini}\gtrsim 15$. 

\begin{figure}
	\centering
	\includegraphics[width=1\hsize]{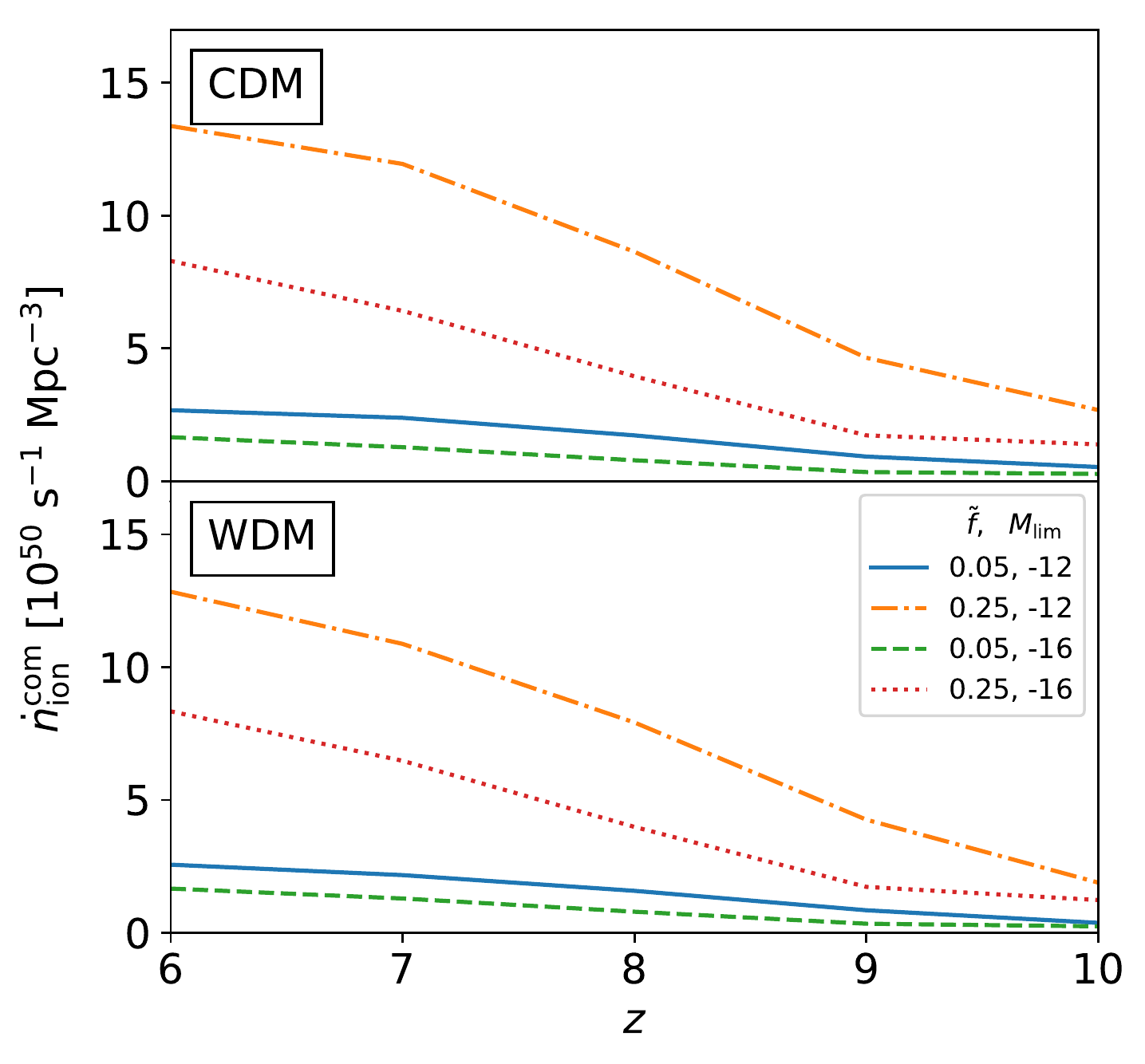}
	\caption{\label{fig:n_ion} The comoving ionizing emissivity $\dot{n}^{\rm com}_{\rm ion}(z)$ interpolated over the redshift interval $6\le z\le10$ for the CDM model (top panel) and WDM model (bottom panel) respectively. The various lines correspond to different combination of values of $\tilde{f}$ and $M_{\rm lim}$ quoted in the legend.}
\end{figure}

In Fig.~\ref{fig:n_ion} we plot the interpolated values of $\dot{n}^{\rm com}_{\rm ion}(z)$ for the CDM (top panel) and WDM (bottom) scenarios and different combination of values of $\tilde{f}$ and $M_{\rm lim}$ shown in the legend. We may notice that $\dot{n}^{\rm com}_{\rm ion}$ is a decreasing function of redshift, which is a direct consequence of the fact that the LF decreases in amplitude at higher redshifts. Overall, this corresponds to injecting fewer ionizing photons in the IGM. In the same line of reasoning, higher values of $\tilde{f}$ or fainter $M_{\rm lim}$ increase the overall amplitude of $\dot{n}^{\rm com}_{\rm ion}$. Differences between DM models are higher when considering fainter $M_{\rm lim}$ since this is the range of luminosities where the LFs differ the most. As an example, in the case $(\tilde{f}, M_{\rm lim}) = (0.25,-12)$, we can clearly see that the ionizing photon emissivity of the CDM model is slightly larger than that of the WDM one. Over the redshift range considered, this is consistent with the fact that the LF of the WDM model is suppressed compared to the CDM case at faint UV-magnitudes as shown in Fig.~\ref{fig:LF}. For $\tilde{f}=0.05$, such differences are still present, though reduced in amplitude by a factor of $5$ and thus not distinguishable by visual inspection of the corresponding curves in Fig.~\ref{fig:n_ion}.

\section{Dark Matter Model Imprints on Cosmic Reionization}\label{resultsI}

\subsection{Optical Depth}

\begin{figure*}
	\centering
	\includegraphics[width=1\hsize]{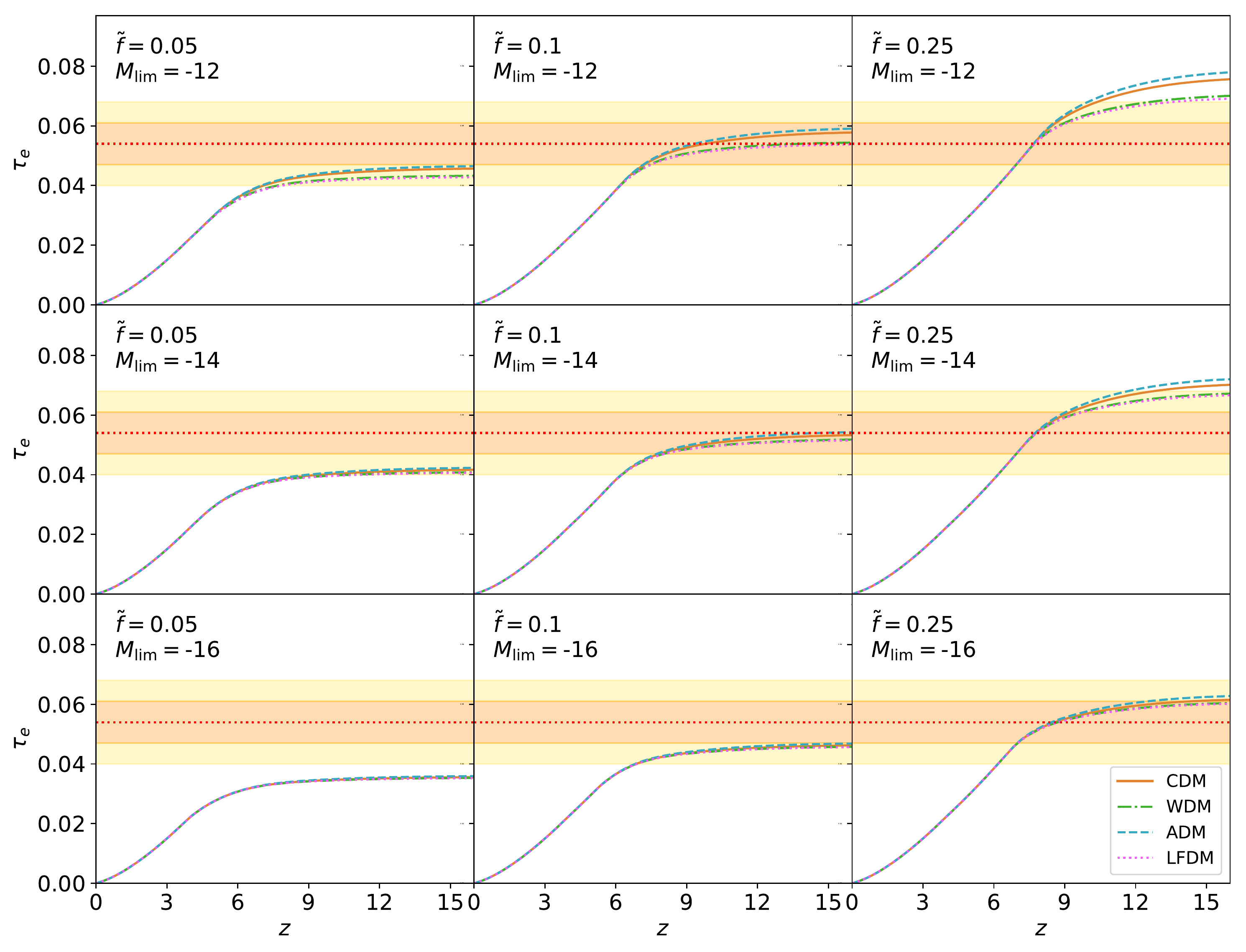}
	\caption{\label{fig:tau_all} Integrated Thomson scattering optical depth $\tau_e(z)$ for the CDM model (orange solid lines), WDM (green dot-dashed lines), ADM (blue dashed lines) and LFDM (magenta dotted lines) for values of $\tilde{f} = 0.05, 0.1,0.25$  (panels from left to right) and $M_{\rm lim} = -12, -14,-16$ (panels from top to bottom). The horizontal dotted line and the shaded yellow area correspond to the mean, $1\sigma$ error (dark area) and $2\sigma$ error (light area) of the integrated optical depth from {\it Planck} \cite{PlanckCosmo2018}.}
\end{figure*}

Free electrons in the reionized IGM interact with CMB photons through Thomson scattering, thus leaving an imprint of the cosmic reionization on the CMB temperature and polarization anisotropies. This integrated effect along the line of sight is quantified by the Thomson scattering optical depth,
\begin{equation}\label{eq:tau}
\tau_e(z) = \int_0^z dz' \frac{c (1+z')^2}{H(z')} Q_{\rm HII}(z') \, \sigma_{\rm T} \, \bar{n}_{\rm H} \, \left(1 + \eta \frac{Y}{4X}\right),
\end{equation}
where $H(z)$ is the Hubble rate, $\sigma_{\rm T}$ is the Thomson cross section, $c$ is the speed of light, $X = 0.75$ and $Y=0.25$ are the hydrogen and helium primordial mass fractions, where $\eta=2$ for $z\leq4$ and $\eta=1$ at higher redshift such as to model the helium double reionization occurring at late times.

We compute the redshift evolution of the optical depth from Eq.~(\ref{eq:tau}) for a given DM scenario assuming different combinations of the reionization model parameters $\tilde{f}$ and $M_{\rm lim}$. We plot the results in Fig.~\ref{fig:tau_all}, where the horizontal dotted line, the dark yellow shaded area and the light yellow shaded area correspond to the mean, $1\sigma$ and $2\sigma$ error of the integrated optical depth from {\it Planck} cosmological data analysis, $\tau_e = 0.054 \pm 0.007$ \cite{PlanckCosmo2018}. The panels from left to right show $\tau_e(z)$ for increasing values of $\tilde{f}$ at fixed $M_{\rm lim}$, while panels from top to bottom correspond to brighter limiting UV-magnitude $M_{\rm lim}$ at constant $\tilde{f}$. In each panel the different lines correspond to the predictions of the CDM (orange solid line), WDM (green dot-dashed line), ADM (blue dashed line) and LFDM (magenta dotted line) models respectively. As expected increasing values of $\tilde{f}$ or fainter $M_{\rm lim}$ lead to higher values of the optical depth. Also, we can see that for a given pair of values of $\tilde{f}$ and $M_{\rm lim}$ the optical depth is systematically lower for WDM and LFDM than CDM and ADM respectively. This is a direct consequence of the differences of the faint-end slope of the corresponding galaxy LFs shown in Fig.~\ref{fig:LF}. We may also notice that at fixed values of $M_{\rm lim}$, the differences among the DM model predictions of $\tau_e(z)$ increase for increasing values of $\tilde{f}$. We find that certain reionization model parameter configurations are excluded by the {\it Planck} bounds independent of the DM model considered. This is the case of the pairs $(\tilde{f}, M_{\rm lim})=(0.25,-12)$, $(0.05,-14)$, $(0.25,-14)$, $(0.05,-16)$ for which the predicted $\tau_e(z)$ either overshoot or underestimate the {\it Planck} constraints. The remaining configurations show the degeneracy between the DM scenario and the reionization model parameters. In fact, for each DM model it is possible to find a combination of values of $\tilde{f}$ and $M_{\rm lim}$ resulting in the same $\tau_e(z)$ while still satisfying the {\it Planck} bounds. In principle, the trends shown in Fig.~\ref{fig:tau_all} suggest that it should be possible to find a pair of values of $\tilde{f}$ and $M_{\rm lim}$ such that the differences between the different DM models are larger, while still in agreement with the {\it Planck} constraints. As we will see next, this is excluded by direct measurements of the volume filling fraction of ionized hydrogen, which contributes to reducing the degeneracy among the reionization model parameters.

\subsection{Redshift Evolution of Hydrogen Ionization Fraction}

\begin{figure*}
	\centering
	\includegraphics[width=1\hsize]{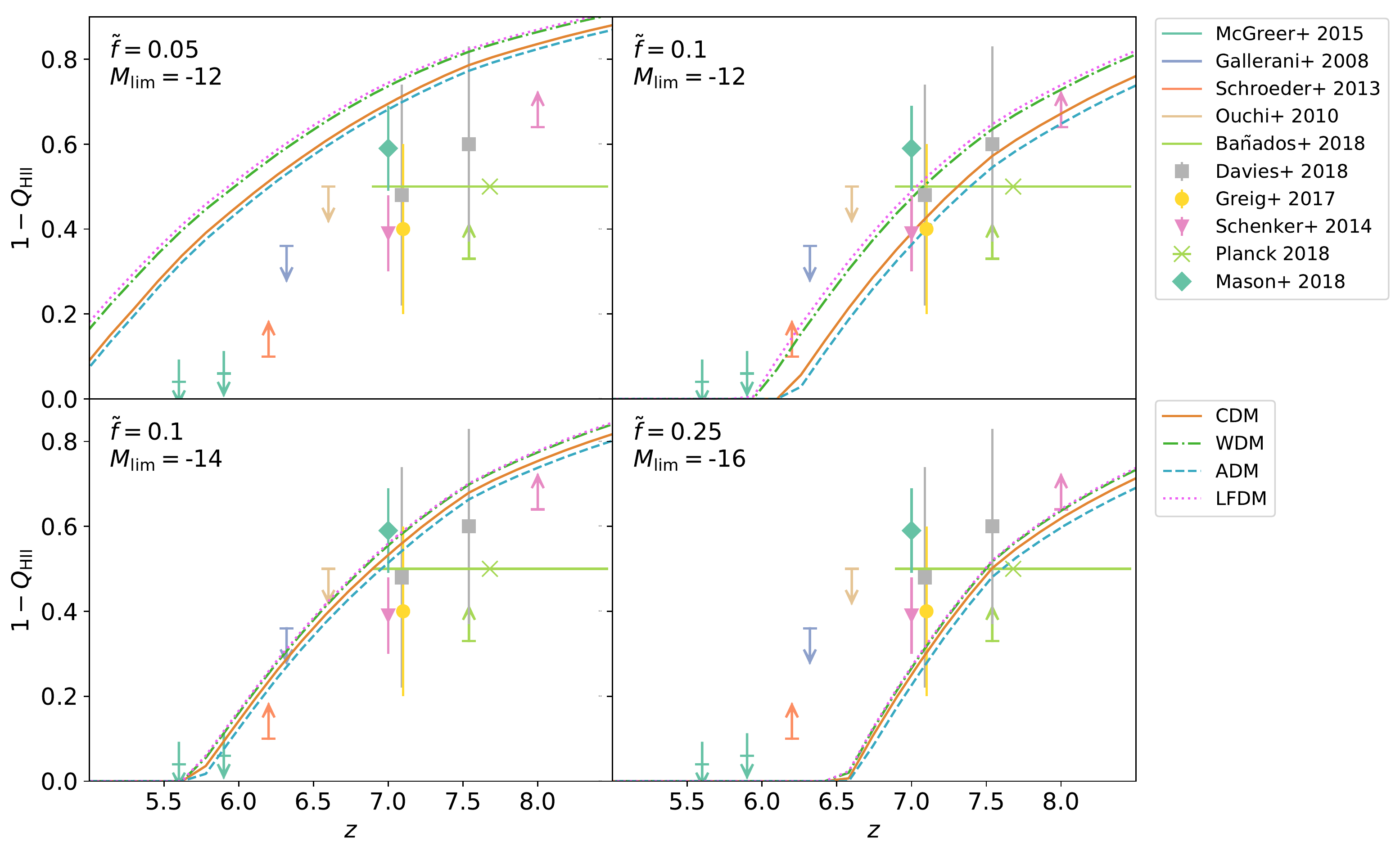}
	\caption{\label{fig:Q_all} Redshift evolution of the neutral hydrogen fraction $1 - Q_{\rm HII}(z)$ in CDM (orange solid lines), WDM (green dot-dashed lines), ADM (blue dashed lines) and LFDM (magenta dotted lines) for different combination of values of $\tilde{f}$ and $M_{\rm lim}$. The different data points and arrows correspond to detections and observational limits from various works in the literature (see text for the corresponding references).}
\end{figure*}

We compute the redshift evolution of the volume filling fraction of ionized hydrogen $Q_{\rm HII}$ by solving Eq.~(\ref{QHII}) for the different DM models. For conciseness, we only focus on a reduced set of values of $\tilde{f}$ and $M_{\rm lim}$ which capture the most relevant model parameter dependencies. Furthermore, to compare with observational constraints we focus on the evolution of the neutral hydrogen fraction, $1 - Q_{\rm HII}(z)$, which we plot in Fig.~\ref{fig:Q_all}. In particular, we show a compilation of measurements of the neutral hydrogen fraction obtained from the analysis of the damping wing of absorption profiles of quasars \cite{Schroeder2013,Greig2017,Davies2018,Banados2018} and from their spectral features \cite{Gallerani2008,McGreer2015}. Other observational constraints are inferred from studies of Lyman-$\alpha$ emitter galaxies, through their redshift distribution \cite{Schenker2014} and clustering properties \cite{Ouchi2010}, and the detection/non-detection of Lyman-$\alpha$ emission in young star-forming galaxies at high redshift (Lyman Break galaxies) \cite{Mason2018}. For illustrative purposes we also show a data-point indicating the redshift at which $Q_{\rm HII} = 0.5$ as inferred from the latest {\it Planck} data analysis \cite{PlanckCosmo2018}.

We may notice that for fixed values of $\tilde{f}$ and $M_{\rm lim}$, the neutral hydrogen fraction decays earlier in ADM and CDM than in WDM and LFDM. Moreover, the differences among the different DM model predictions increase for fainter limiting UV-magnitudes, consistent with expectations from the differences of the faint-end slope of the corresponding galaxy LFs. We can also see that for decreasing values of $\tilde{f}$, the curves are shifted towards lower redshifts, thus leading to a delayed reionization process. It is worth noticing that independent of the DM model certain combinations of the reionization model parameters are excluded by measurements of the neutral hydrogen fraction. Hence, as already mentioned, these measurements contribute to reducing the reionization model parameter degeneracy, which we will discuss in more detail in Section~\ref{MCMC}. 

\subsection{Duration of the Cosmic Reionization}

\begin{figure*}
	\centering
	\includegraphics[width=0.8\hsize]{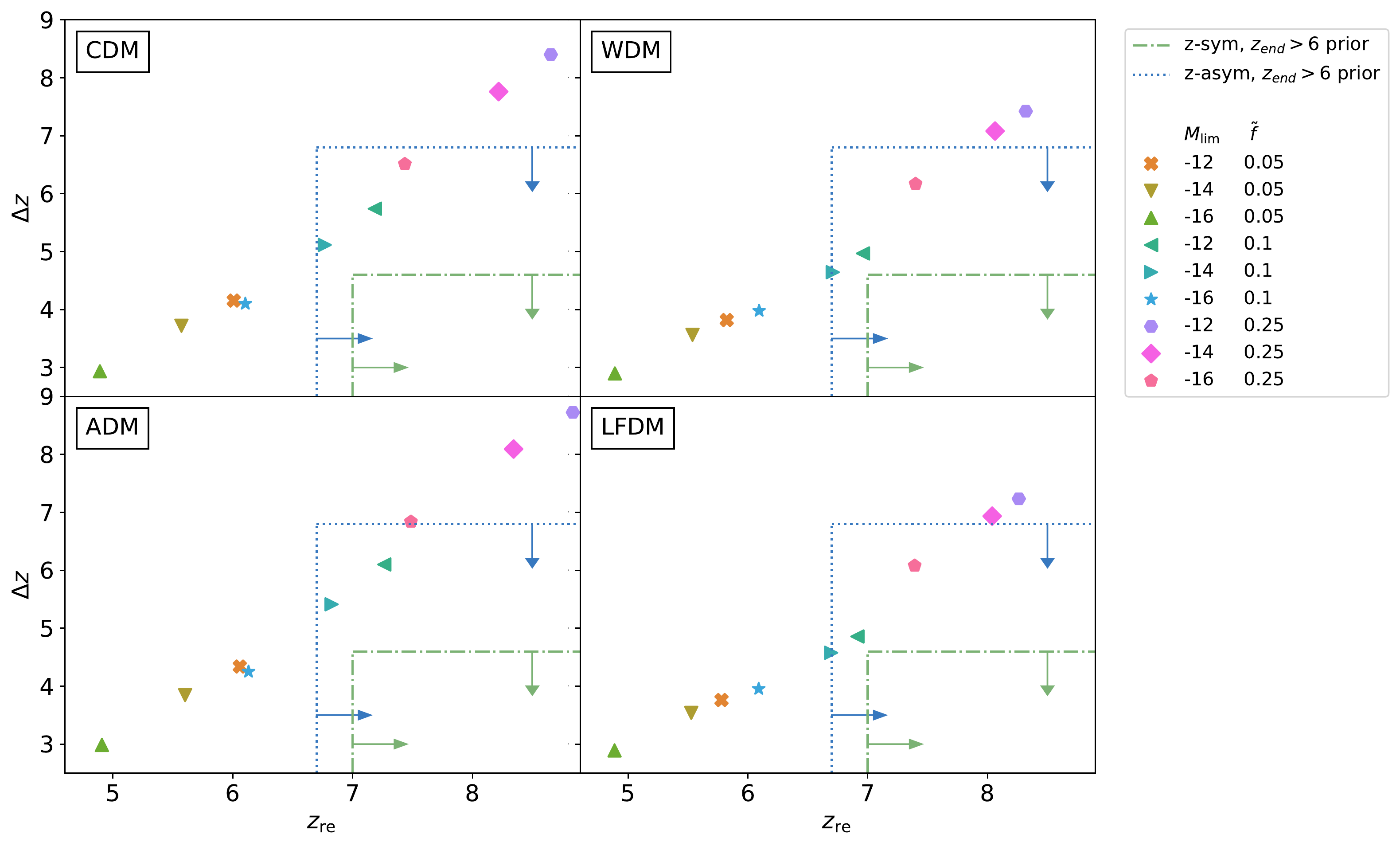}
	\caption{\label{fig:planck} Reionization redshift $z_{\rm re}$ and duration $\Delta z = z_{0.10} - z_{0.99}$ for the CDM (top left panel), WDM (top right panel), ADM (bottom left panel) and LFDM (bottom right panel) models with different values of $\tilde{f}$ and $M_{\rm lim}$ specified by the markers in the legend. The rectangular areas in the panels delimit the $2\sigma$ confidence region for a $z$-symmetric reionization model (green dot-dashed line) and $z$-asymmetric one (blue dotted line) from \cite{PlanckReion2016}.}
\vspace{1cm}
	\includegraphics[width=0.8\hsize]{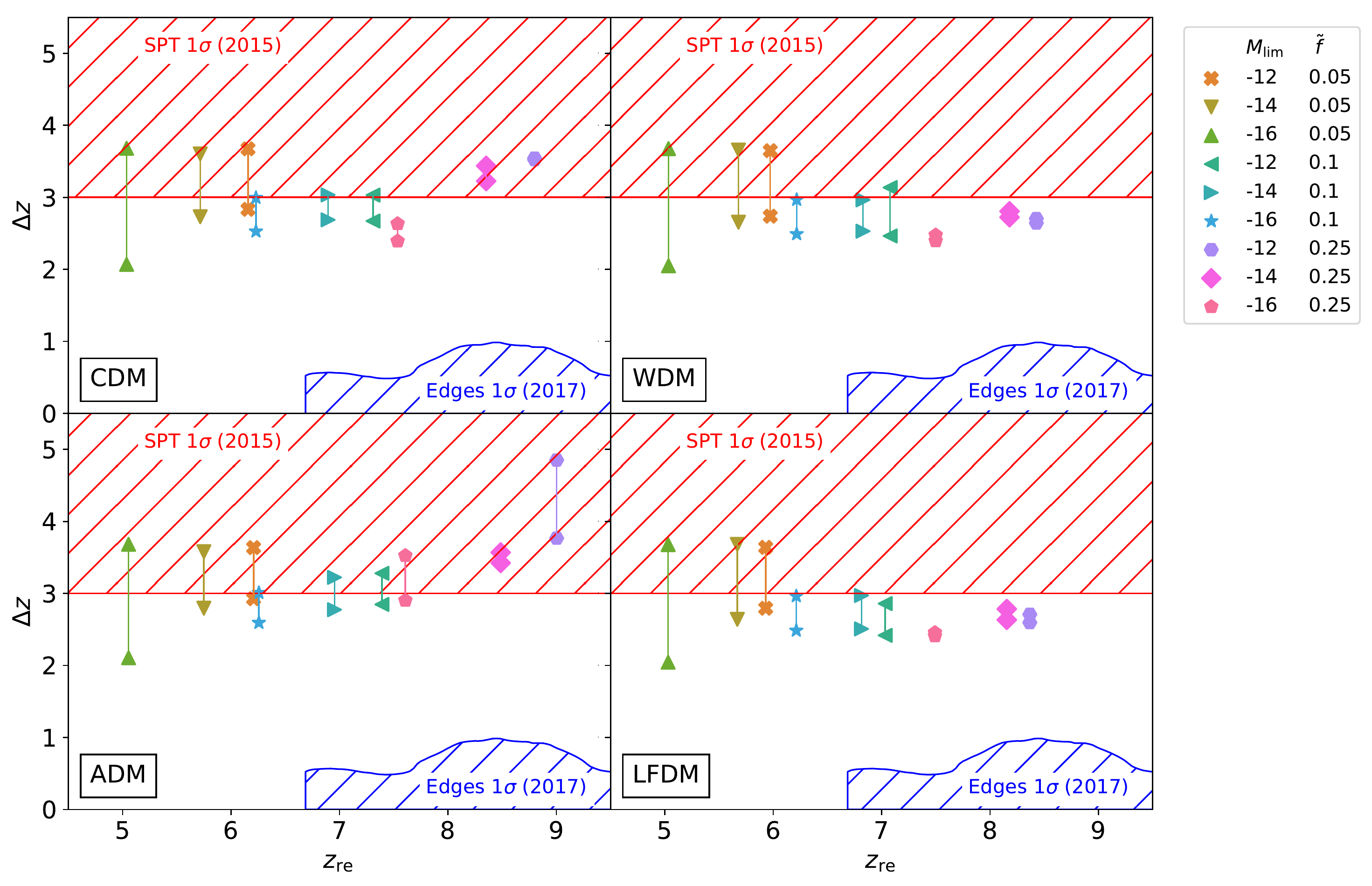}
	\caption{\label{fig:edges} Reionization redshift $z_{\rm re}$ and duration $\Delta z$ for the CDM (top left panel), WDM (top right panel), ADM (bottom left panel) and LFDM (bottom right panel) models with different values of $\tilde{f}$ and $M_{\rm lim}$ specified by the markers in the legend. For each pair of markers, the bottom marker indicates the redshift duration computed as $\Delta z = z_{0.20} - z_{0.99}$ following the definition assumed in \cite{George2015}, whereas the top one has been calculated as $\Delta z = - \left( d Q_{\rm HII} / d z\right)^{-1}|_{Q_{\rm HII}=0.5}$ following the definition of \cite{Monsalve2017}. The first is the definition used to derive the SPT excluded region (top red hatched region), while the latter has been used to infer the EDGES one (bottom blue hatched region).}
\end{figure*}

The redshift evolution of the cosmic reionization history can be characterized in terms of a central redshift $z_{\rm re}$ defined as the redshift at which half of the hydrogen in the IGM is ionized, $z_{\rm re}= z(Q_{\rm HII}=0.5)$, and the duration of the process $\Delta{z}$. 

The {\it Planck} collaboration has performed a thorough analysis of the imprint of the reionization history on the CMB temperature and polarization power spectra from the intermediate data release \cite{PlanckReion2016}. In particular, they have inferred constraints on $z_{\rm re}$ and $\Delta{z}\equiv z_{0.10}-z_{0.99}$ with $z_x\equiv z(Q_{\rm HII}=x)$ assuming two different redshift parametrizations of the ionization fraction. These consist of a $z$-symmetric model, which parametrizes a step-like transition in redshift; a $z$-asymmetric model corresponding to a power law behavior above the redshift at which reionization is completed and a constant value at lower redshifts. It is worth remarking that the latter model provides a better description to the $Q_{\rm HII}(z)$ predictions shown in Fig.~\ref{fig:Q_all}.

In Fig.~\ref{fig:planck} we plot the values of $z_{\rm re}$ and $\Delta{z}$ for the different DM scenarios and different combinations of $\tilde{f}$ and $M_{\rm lim}$. The constraints from {\it Planck} \cite{PlanckReion2016} are delimited by the rectangular areas in the lower bottom part of each panel. These denote the $2\sigma$ confidence regions inferred assuming the $z$-asymmetric (blue dotted line) and $z$-symmetric cases (green dot-dashed line), with prior $z_{\rm end}=6$ on the redshift at which reionization is completed. We may noticed that for each DM scenario the sets of values of $\tilde{f}$ and $M_{\rm lim}$ predicting values of $z_{\rm re}$ and $\Delta{z}$ in agreement with the {\it Planck} limits, are also those which are consistent with the $1\sigma$ constraint on $\tau_e$ (Fig.~\ref{fig:tau_all}).

Measurements of the kinetic Sunyaev-Zel'dovich (kSZ) power spectrum can independently constrain the duration of the reionization process. Ionizing bubbles of gas forming around the source of cosmic reionization expand and merge till the universe is fully ionized. CMB photons scattering on these moving ionized gas clouds, generate a kSZ imprint on the CMB whose amplitude depends on the duration of the cosmic reionization. Measurements of the kSZ power spectrum from the South Pole Telescope have provided constraints on the duration of the cosmic reionization, defined as $\Delta z \equiv z_{0.20} - z_{0.99}$ (see \cite{George2015}). These constraints depend on the specifics of the patchy reionization model considered by the authors of \cite{George2015} to interpret the kSZ signal. Nonetheless, these measurements are worth mentioning, as they provide additional constraints on the cosmic reionization models to be integrated in future data analysis.

In Fig.~\ref{fig:edges} we plot the region of $\Delta{z}$ values excluded at $1\sigma$ by the SPT measurements (red hatched area). We also plot the region excluded by the analysis of \cite{Monsalve2017} (blue hatched area) based on the limit on the global 21-cm signal derived by the high-band data of the EDGES experiment. Different from \cite{George2015}, the authors of \cite{Monsalve2017} define the duration $\Delta z \equiv - \left( d Q_{\rm HII} / d z\right)^{-1}|_{Q_{\rm HII}=0.5}$. In Fig.~\ref{fig:edges} we plot the corresponding values of $\Delta{z}$ for the different DM models and different combinations of values of $\tilde{f}$ and $M_{\rm lim}$. In each panel the predictions are shown as a pair of linked markers, the top one indicating the value obtained using the definition of \cite{Monsalve2017} and the bottom obtained from the definition of \cite{George2015}. 
In the former case the estimated values of $\Delta{z}$ are well within the region consistent with the EDGES bound. The predictions of the WDM and LFDM models are in agreement with the SPT limit, whereas in the case of the CDM and ADM models the configurations with $(\tilde{f}, M_{\rm lim}) = (0.25, -12)$ and $(0.25, -14)$ seem to be in tension with SPT result. However, given the assumptions made in the derivation of this SPT bound such a discrepancy is not significant. Indeed, relaxing the SPT limit to the $2\sigma$ confidence level, the bound increases to $\Delta{z} = 5.4$, thus removing any tension.

\section{Constraints on Reionization Model Parameters}\label{MCMC}

We perform a likelihood analysis to derive constraints on the reionization model parameters for a given DM scenario using the {\it Planck} determination of the integrated optical depth \cite{PlanckCosmo2018} in combination with high redshift measurements of the neutral hydrogen fraction from \cite{Schenker2014,Greig2017,Davies2018}. To be as conservative as possible we do not include constraints on the median redshift $z_{\rm re}$ and duration of reionization $\Delta z$ discussed in the previous section. In fact, these bounds are not obtained directly from observations, rather they are inferred from data analyses which assume a reionization history model whose predictions may differ from those that we have considered here. Furthermore, as shown in Figs.~\ref{fig:planck}-\ref{fig:edges}, the constraints on $z_{\rm re}$ and $\Delta z$ point to the same combinations of ($\tilde{f}, M_{\rm lim}$) probed by $\tau_e$ and $Q_{\rm HII}(z)$. Thus, given the large uncertainties, these may not carry sufficiently accurate independent information to break reionization model parameter degeneracies. 
In the previous section, we have shown that the reionization histories predicted in the case of ADM and LFDM models closely follow those of CDM and WDM respectively, thus without loss of generality we limit our analysis to the latter models. We sample a two-dimensional parameter space with $\tilde{f}\in [0.05,0.25]$ and $M_{\rm lim}\in [-16,-12]$. 

\begin{figure}
	\centering
	\includegraphics[width=1\hsize]{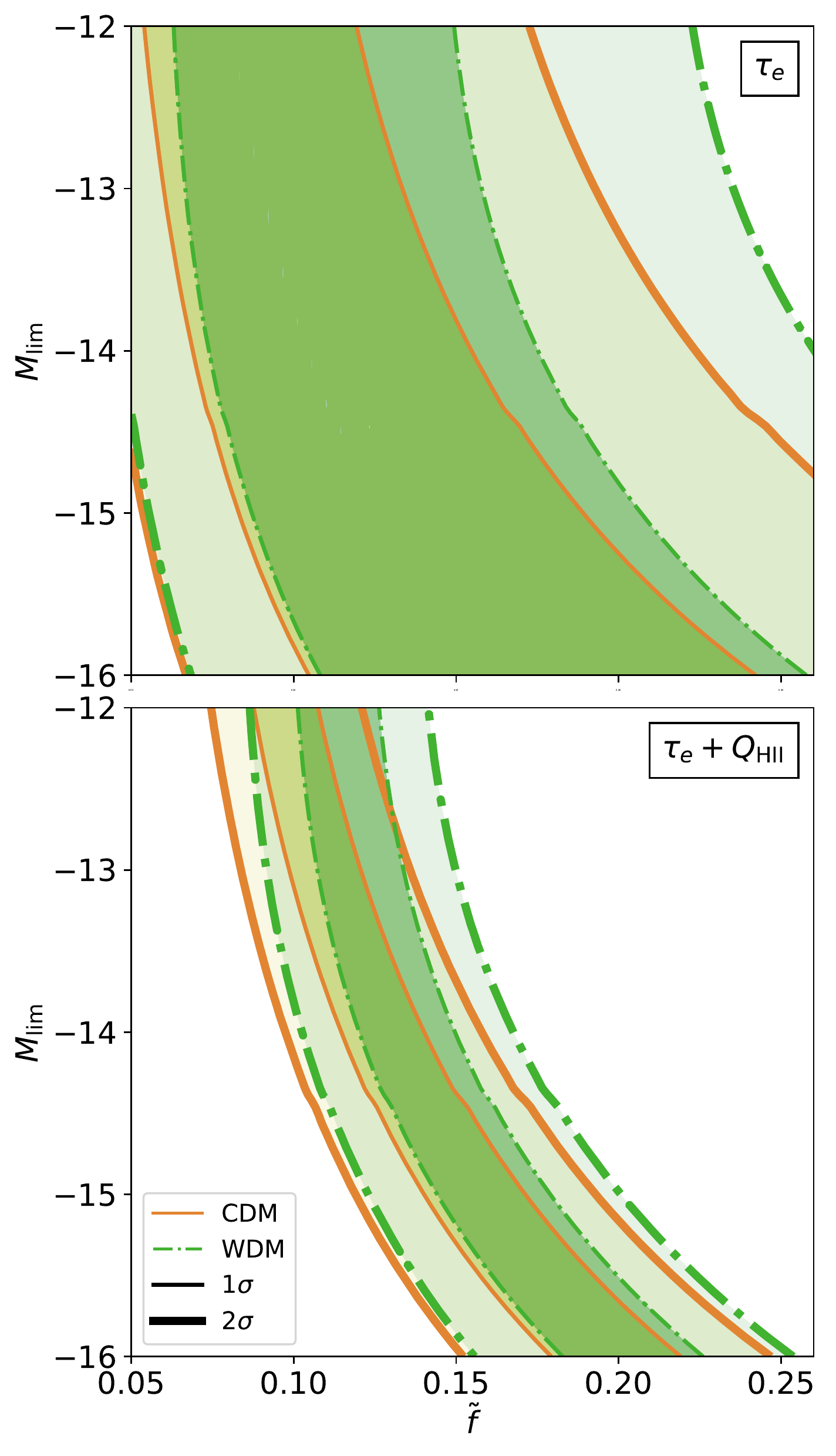}
	\caption{\label{fig:space} Confidence regions at $1\sigma$ (dark shaded area) and $2\sigma$ (light shaded area) in the $\tilde{f}$-$M_{\rm lim}$ plane for the CDM (orange solid lines) and WDM (green dash-dotted lines) respectively. The top panel shows the constraints inferred using the {\it Planck} determination of the optical depth $\tau_e$ \cite{PlanckCosmo2018}, while the bottom panel shows the constraints obtained in combination with measurements of $Q_{\rm HII}$ from \cite{Greig2017,Schenker2014,Davies2018}.}
\end{figure}

In the top panel of Fig.~\ref{fig:space} we plot confidence contours at $1\sigma$ (dark shaded area) and $2\sigma$ (light shaded area) in the $\tilde{f}$-$M_{\rm lim}$ plane as inferred using the {\it Planck} determination of $\tau_e$ for the CDM (orange solid lines) and WDM (green dash-dotted lines) models respectively. We can see that the constraints are very loose due to the degeneracy between $\tilde{f}$ and $M_{\rm lim}$. The confidence contours of the two DM scenarios largely overlap, though we may notice that for faint limiting UV-magnitudes ($M_{\rm lim}\gtrsim -15$) and large values of the effective escape fraction ($\tilde{f}\gtrsim 0.20$) the CDM constraints are slightly tighter than those of the WDM model. This is because in such a region of the parameter space the CDM prediction overshoots the value of $\tau_e$ from {\it Planck} (see Fig.~\ref{fig:tau_all}). In contrast, in the same region the WDM model systematically predicts a value of $\tau_e$ lower than CDM, thus in better agreement with the measured value of $\tau_e$. 

In the bottom panel of Fig.~\ref{fig:space} we plot the confidence contours inferred from the combined analysis of $\tau_e$ and $Q_{\rm HII}(z)$ estimates from \cite{Schenker2014,Greig2017,Davies2018}. Not surprisingly, we infer much tighter constraints in the $\tilde{f}$-$M_{\rm lim}$ plane, though the degeneracy between $\tilde{f}$ and $M_{\rm lim}$ still persists. As expected, the contours from CDM and WDM slightly differ at faint limiting UV-magnitudes and for low values of the effective escape fraction, which is a region of the reionization model parameter space where the predictions of CDM and WDM differ the most. In both cases the degeneracy lines seem to tend toward an asymptotic trend for low values of $M_{\rm lim}$. In particular, for $M_{\rm lim}\gtrsim -13$, then the effective escape fraction is bound to be in the rage $0.07\lesssim \tilde{f}\lesssim 0.15$ at $2\sigma$ independent of the DM scenario considered\footnote{At $z\approx 10$ a limiting UV-magnitude $M_{\rm lim}\gtrsim -13$ corresponds approximately to a minimum star forming halo mass $M_h^{\rm min}\lesssim 5\cdot 10^9\,M_{\odot}\,h^{-1}$ in the CDM case, while $M_h^{\rm min}\lesssim 7\cdot 10^8\,M_{\odot}\,h^{-1}$ in the WDM model.}.

It is encouraging that the $\tilde{f}$-$M_{\rm lim}$ confidence regions shown in the bottom panel of Fig.~\ref{fig:space} overlap at a $2\sigma$ level with that of the analysis by Kakiichi et al. \cite{Kakiichi2018} (see their Fig. 9). The authors of \cite{Kakiichi2018} infer their constraints correlating spatial positions of star-forming galaxies at $z\sim 6$ with the Lyman-$\alpha$ forest seen in the spectrum of a background quasar, thus providing a different and complementary study of what we propose here.

The bounds shown in Fig.~\ref{fig:space} indicate that current bounds on the cosmic reionization history are largely insensitive to the specifics of the underlying DM scenario. Overall, they suggests that the observed abundance of faint galaxies at high redshift can account for independent tests of the cosmic reionization history from CMB and neutral hydrogen fraction measurements provided that the effective escape fraction and the limiting UV-magnitudes are within the confidence regions shown in Fig.~\ref{fig:space}. The availability of more precise measurements of the neutral hydrogen fraction may further narrow these constraints. However, breaking the internal parameter degeneracy require additional independent proxies of the cosmic reionization, which depending upon the region of the parameter space may be sensitive to the ultimate nature of DM.

\section{Redshift Evolution of Escape Fraction}\label{resultsII}

\begin{figure}
	\centering
	\includegraphics[width=1\hsize]{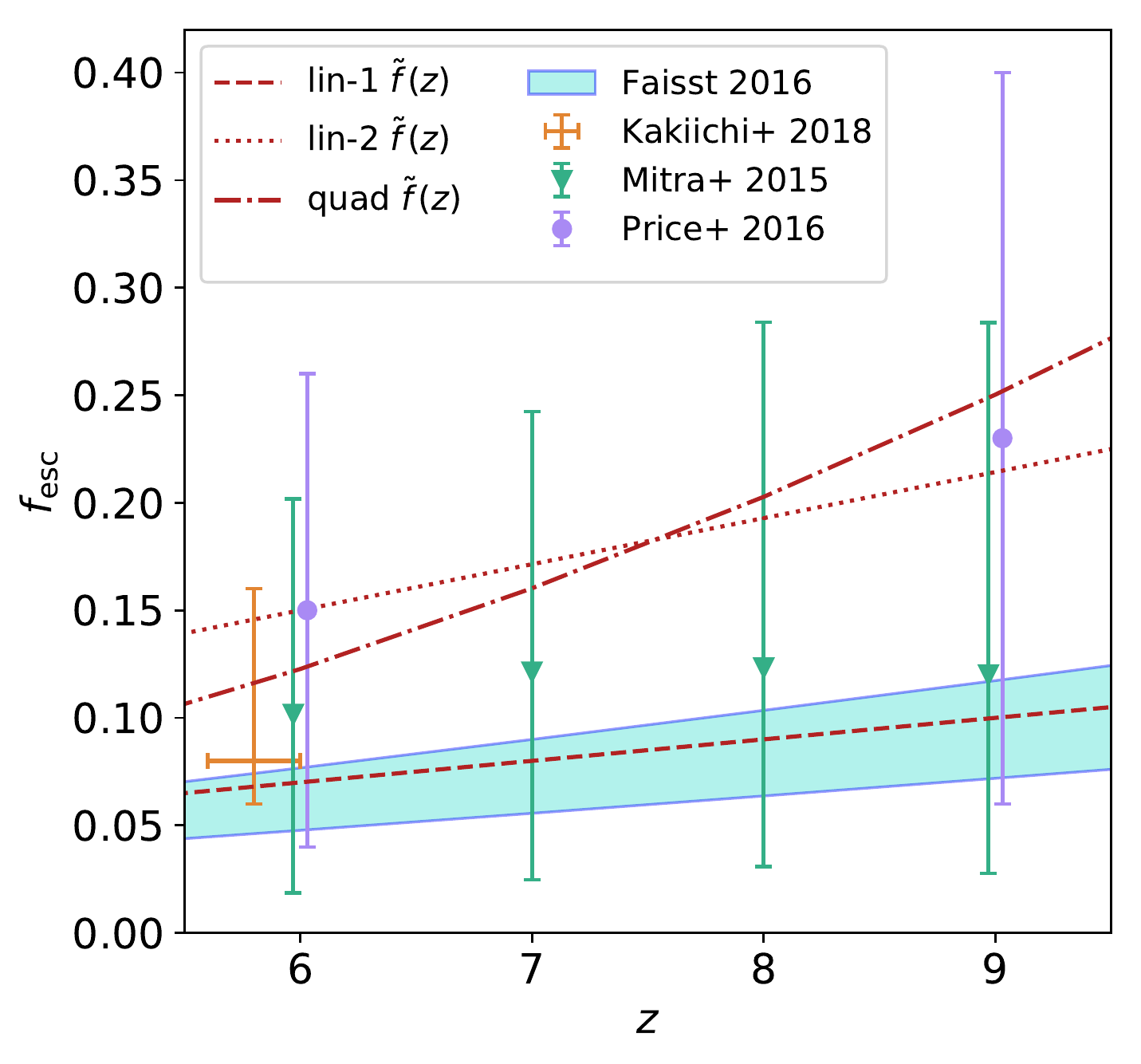}
	\caption{\label{fig:fz} Measurements of the escape fraction $f_{\rm esc}$ at different redshifts from \cite{Mitra2015,Price2016,Faisst2016,Kakiichi2018}. We also plot linear parametrization models ``lin-1'' (red dashed line) and ``lin-2'' (red dotted line) and the quadratic model ``quad'' (red dot-dashed line).}
\end{figure}

\begin{figure*}
	\centering
	\includegraphics[width=0.9\hsize]{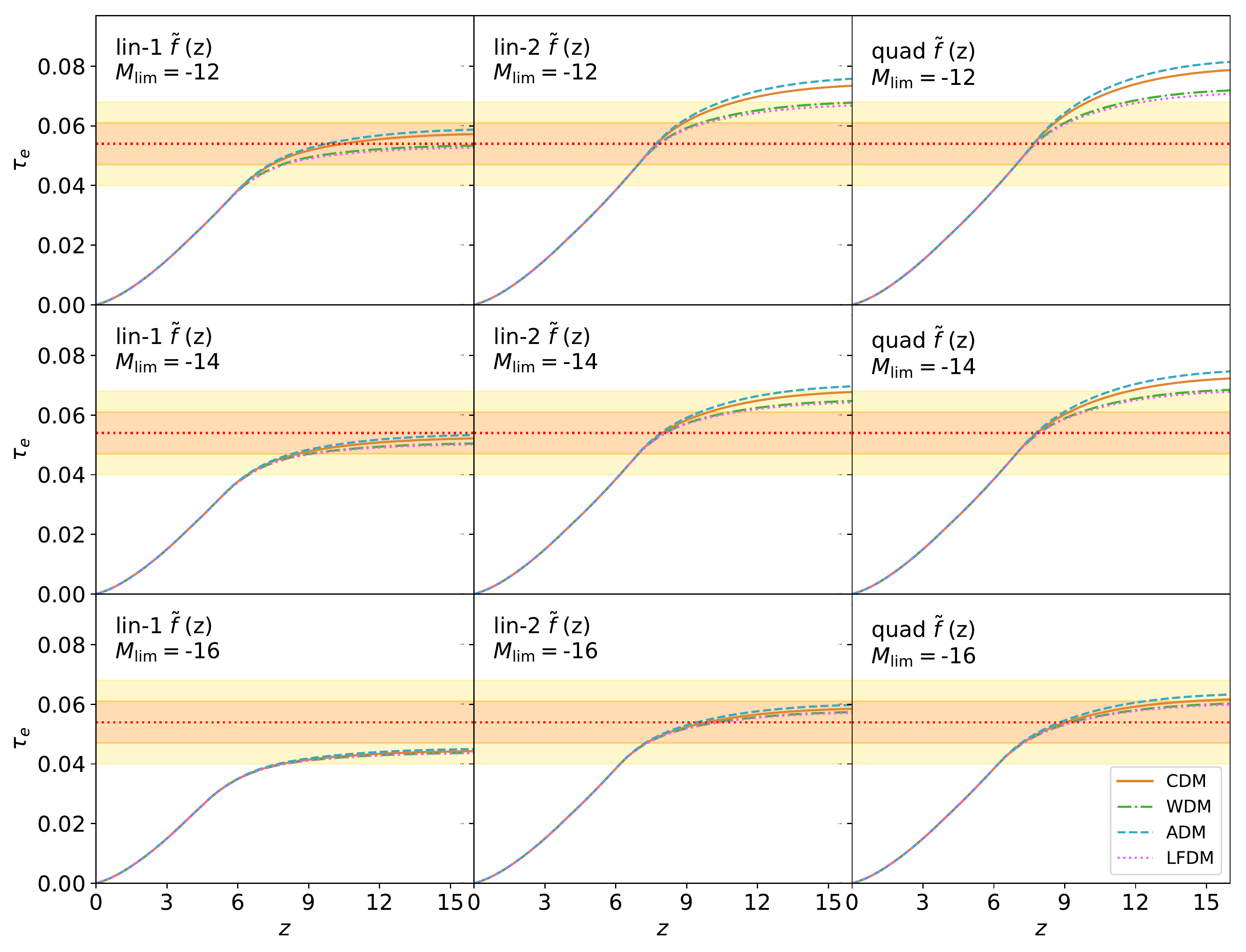}
	\caption{\label{fig:tau_all_fz} Integrated Thomson scattering optical depth $\tau_e(z)$ as in Fig.~\ref{fig:tau_all} for the ``lin-1'', ``lin-2'' and ``quad'' redshift parametrization of the escape fraction (panels from left to right) and $M_{\rm lim} = -12, -14,-16$ (panels from top to bottom). The horizontal dotted line and the shaded yellow area correspond to the mean and $1\sigma$ (dark) and $2\sigma$ (light) error of the integrated optical depth from {\it Planck} \cite{PlanckCosmo2018}.}
\end{figure*}

In the previous sections we have discussed the cosmic reionization history of different DM scenarios assuming a redshift independent escape fraction. Here, we intend to investigate to which extent the trends derived in Section~\ref{resultsI} remain valid when we assume a monotonic evolution of $\tilde{f}$ consistent with state-of-art measurements of the escape fraction. To this purpose we test a linear model with $\tilde{f}^{\rm lin}(z) = \tilde{f}_{z=6} (1+z)/7$ for two different values of $\tilde{f}_{z=6}=0.07$ (lin-1) and $0.15$ (lin-2) respectively, and a quadratic model $\tilde{f}^{\rm quad}(z) = a (1+z)/1000 + b (1+z)^2/100$ with $a = 0.03$ and $b=0.25$ (quad). These are shown in Fig.~\ref{fig:fz} against a compilation of measurements from \cite{Mitra2015,Price2016,Faisst2016,Kakiichi2018}. We can see that a constant escape fraction is consistent with currently available data, as well as an increasing function of redshift. The latter is qualitatively in agreement with the hypothesis that early galaxies, being more pristine and dust-poor, allow for a larger escape fraction of ionizing photons since they consist of very strong and luminous metal-poor stars (although this scenario has been recently challenged by the findings of \cite{Laporte2017}). 

\begin{figure*}
	\centering
	\includegraphics[width=0.8\hsize]{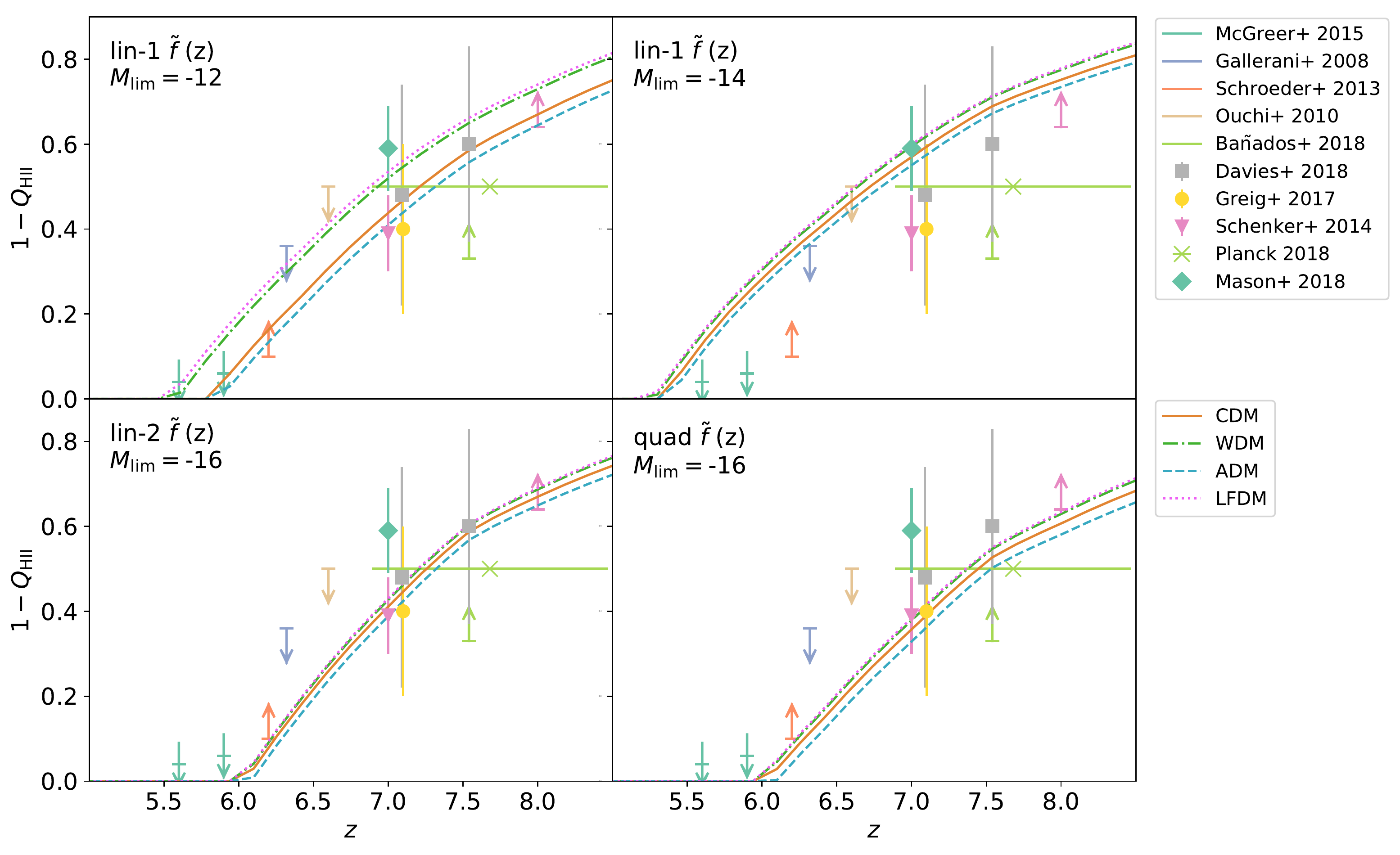}
	\caption{\label{fig:Q_fz} Redshift evolution of the neutral hydrogen fraction $1 - Q_{\rm HII}(z)$ as in Fig.~\ref{fig:Q_all} for escape fraction models ``lin-1'' with $M_{\rm lim}=-12$, ``lin-1'' with $M_{\rm lim}=-14$, ``lin-2'' with $M_{\rm lim}=-16$ and ``quad'' with $M_{\rm lim}=-16$ respectively.} 
	\vspace{1cm}
	\includegraphics[width=0.8\hsize]{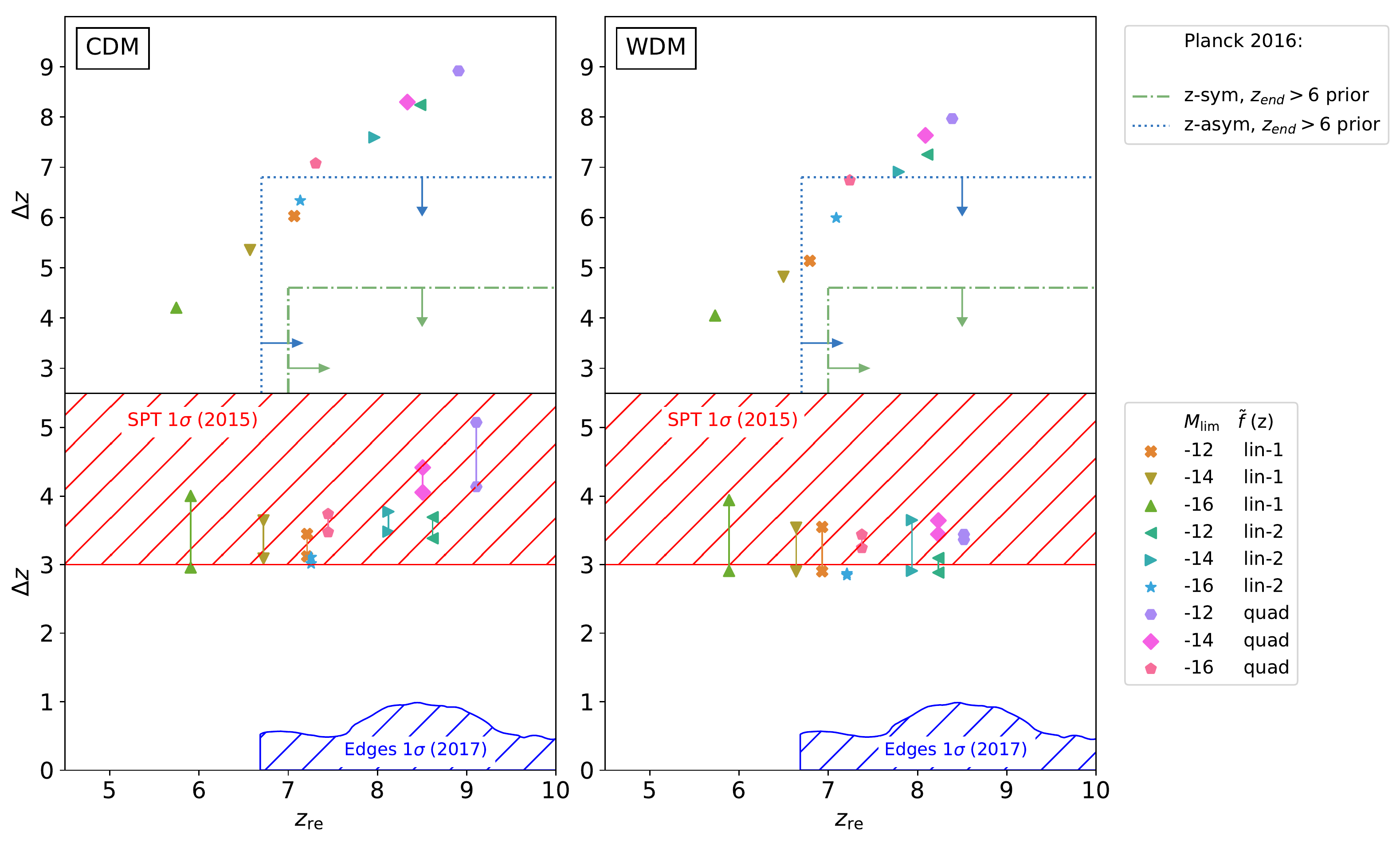}
	\caption{\label{fig:timescale_fz} Reionization redshift $z_{\rm re}$ and its duration $\Delta z$ as in Fig.~\ref{fig:edges} for the different $\tilde{f}(z)$ models and $M_{\rm lim}$ values as in Fig.~\ref{fig:tau_all_fz}. In the top panels we show the observational limits for a $z$-symmetric reionization model (dot-dashed line) and $z$-asymmetric one (dotted line) from \cite{PlanckReion2016}, while in the bottom panels we show the excluded region  from SPT \cite{George2015} and from the high-band data of the EDGES experiment \cite{Monsalve2017}.}   
\end{figure*}

In Fig.~\ref{fig:tau_all_fz} we show the optical depth $\tau_e(z)$ of each DM scenario predicted by the $\tilde{f}(z)$ toy models  (panels from left to right) and values of $M_{\rm lim} = -12, -14$ and $-16$ (panels from top to bottom), against the measurement reported by {\it Planck} (horizontal dashed line and dark and light shaded areas for the $1\sigma$ and $2\sigma$ errors). We can see trends which are similar to those shown in Fig.~\ref{fig:tau_all}. Overall, the monotonic increase of the escape fraction induce slightly larger differences between the different DM models at fixed values of $M_{\rm lim}$, with such differences becoming larger for fainter limiting UV-magnitudes. We can see that certain reionization model assumptions are excluded by the {\it Planck} constraints on $\tau_e$. The ``lin-1'' model is agreement with {\it Planck} for $M_{\rm lim}=-12$ and $-14$, while ``lin-2'' and ``quad'' for $M_{\rm lim}=-16$. In Fig.~\ref{fig:Q_fz}, we plot $1-Q_{\rm HII}(z)$ for these reionization models. These are also consistent with measurements of the neutral hydrogen fraction, though differences among the DM scenarios are indistinguishable and well within the uncertainties. Finally, we plot in Fig.~\ref{fig:timescale_fz} the predicted values of $z_{\rm re}$ and duration $\Delta z$ for the CDM (left) and WDM (right) models for all the $(\tilde{f}(z), M_{\rm lim})$ combinations shown in Fig.~\ref{fig:tau_all_fz} against the limits from SPT and EDGES as presented in Fig.~\ref{fig:edges}. Again, we find that the combinations of model assumptions in better agreement with these limits are those consistent with the constraints on $\tau_e$ and  $1-Q_{\rm HII}$ shown in Figs.~\ref{fig:tau_all_fz} and \ref{fig:Q_fz}.

We conclude this section by stressing that the $\tilde{f}(z)$ scenarios studied here are only toy models far from completely assessing the possible range of $z$-dependent escape fraction models. Nevertheless, through this simple analysis we have shown that a redshift dependent escape fraction does not drastically change the degeneracy among reionization model parameters and DM models.

\section{Conclusions}\label{conclu}

In this work we have investigated the cosmic reionization history of different DM scenarios and derived constraints on the key parameters that shape the evolution of the ionization process. In particular, we have focused on CDM and alternative DM models which are characterized by a suppression of the abundance of low mass halos compared to the CDM prediction. In the high-redshift universe, these halos host faint galaxies which are thought to be the source of the cosmic reionization. In recent years, these have been the target of several observational programs which have provided measurements of the faint end of the galaxy luminosity function at high-redshifts. Here, building upon the work of \cite{Corasaniti2017}, we studied the cosmic reionization history of dark matter scenarios which reproduce the observed galaxy luminosity functions at $6\le z\le 10$. Using a commonly adopted reionization model from \cite{Kuhlen2012}, for each DM model we have predicted observables of the cosmic reionization history parametrized in terms of two astrophysical parameters: the minimum UV-magnitude of galaxies contributing to the reionization process ($M_{\rm lim}$) and a redshift-independent effective escape fraction of UV photons reaching the IGM ($\tilde{f}$). We have computed the redshift evolution of the Thomson scattering optical depth of CMB photons $\tau_{e}(z)$ and the comoving ionized fraction $Q_{\rm HII}(z)$ for each DM models and for different combinations of the reionization model parameters. We have shown that the assumed DM models predict very similar reionization histories. Differences among the model predictions increase for increasing limiting UV-magnitudes, though the imprints are degenerate with the value of $M_{\rm lim}$ and $\tilde{f}$. We have also studied the model dependence of the median redshift and duration of the ionizing process which can be probed through measurements of the kSZ power spectrum. 

We have performed a likelihood analysis to infer constraints on $\tilde{f}$ and $M_{\rm lim}$ in the case of the CDM scenario and a WDM model using {\it Planck} measurement of the integrated optical depth $\tau_e$ and estimates of the neutral hydrogen fraction $1-Q_{\rm HII}$ at different redshifts. The results are quite independent of the specifics of the assumed DM models. The constraints inferred from the analysis of the integrated optical depth only are quite large, due to the internal reionization model parameter degeneracies. On the other hand, including the neutral hydrogen fraction data significantly narrow the confidence regions, though the degeneracy between $\tilde{f}$ and $M_{\rm lim}$ persists. Quite interestingly, we find that for faint limiting UV-magnitudes $M_{\rm lim}\gtrsim -13$ the effective escape fraction lies in the interval $0.07\lesssim \tilde{f}\lesssim 0.15$ at $2\sigma$ independent of the DM model. 

We have also investigated the case of a redshift dependent escape fraction and shown that it does not significantly alter the trends obtained assuming a constant escape fraction.

It is worth remarking that the modeling of the luminosity function adopted here remains agnostic with respect to the baryonic processes responsible for the formation of the faintest galaxies in the universe (for a recent and comprehensive review on early galaxy formation, see \cite{Dayal_review}). These baryonic mechanisms must reproduce the relation between host halo mass and hosted galaxy UV-luminosity (and SFR) which we have derived for each DM model from the analysis of the high-redshift galaxy LF measurements, as shown in Fig.~\ref{fig:MUVMH}. The fact that such a relation differs from one DM model to another at faint UV-magnitude implies that these baryonic processes cannot occur identically. Furthermore, since the predictions of the cosmic reionization for $M_{\rm lim}<-13$ depend on an extrapolation below the range of magnitude covered by the galaxy LF data, we cannot exclude a priori the possibility that baryonic physics may alter the halo mass UV-magnitude relation, especially at very high redshift. Following these arguments, new insights would arise from the study of early galaxy formation in non-standard DM scenarios.

Overall, our analysis suggests that due to large observational uncertainties and the unbroken degeneracy of the reionization model parameters, current probes of the cosmic reionization are insensitive to the specifics of DM scenarios characterized by suppressed abundances of low mass halos. Additional independent measurements of the reionization history are indeed necessary to break the reionization parameter degeneracies. Then, depending on the constrained region of the parameter space these measurements might be sensitive to DM model assumptions. 

\acknowledgements

We are grateful to Adam Lidz for having carefully read our manuscript. IPC wants to thank Richard Ellis, Jose O$\tilde{\rm n}$orbe and Niall Jeffrey for stimulating discussions. This work was granted access to the HPC resources of TGCC under the allocation 2016 - 042287 made by GENCI (Grand Equipement National de Calcul Intensif) on the machine Curie. The research leading to these results has received funding from the European Research Council under the European Community Seventh Framework Programme (FP7/2007-2013 Grant Agreement no. 279954). We acknowledge support from the DIM ACAV of the Region Ile-de-France. We acknowledge the use of the open source Python libraries including: SciPy \cite{scipy}, NumPy \cite{numpy} and Matplotlib \cite{matplotlib}. This work made extensive use of the NASA Astrophysics Data System and of the astro-ph preprint archive at arXiv.org.

\end{document}